\documentclass{iopart}
\usepackage{cite}
\usepackage{iopams}
\usepackage{graphicx}
\usepackage{calc}
\usepackage{amssymb}
\usepackage{color}
\usepackage[pdftex,dvipsnames,usenames]{xcolor}
\usepackage[colorlinks=true,urlcolor=blue,citecolor=blue,linkcolor=blue]{hyperref}

\newcommand{\ket}[1]{\left|#1\right\rangle}
\newcommand{\bra}[1]{\left\langle #1\right|}
\renewcommand{\Tr}{\mbox{\rm Tr}}
\newcommand{\normord}[1]{:\mathrel{#1}:}
\newcommand{\Perm}{\mbox{\rm Perm}}
\newcommand{\D}{\rm d}

\begin{document}
\title{Realistic photon-number resolution in generalized Hong-Ou-Mandel experiment}

\author{V.~Ye.~Len$^{1,2}$, M.~M.~Byelova$^{1,2}$, V.~A.~Uzunova$^{3,4}$, and A.~A.~Semenov$^{1,3,5}$}

\address{$^1$Bogolyubov Institute for Theoretical Physics, NAS of Ukraine, Vul.~Metrologichna 14b, 03143 Kyiv, Ukraine}
\address{$^2$Physics Department, Taras Shevchenko National University of Kyiv, Prospect Glushkova 2, 03022 Kyiv, Ukraine}
\address{$^3$Institute of Physics, NAS of Ukraine, Prospect Nauky 46, 03028 Kyiv, Ukraine}
\address{$^4$Institute of Theoretical Physics, Faculty of Physics, University of Warsaw, ul.~Pasteura 5, 02-093 Warszawa, Poland}
\address{$^5$Kyiv Academic University, Blvd. Vernadskogo 36, 03142  Kyiv, Ukraine}
\date{\today}

\begin{abstract}
	We consider realistic photodetection in a generalization of the Hong-Ou-Mandel experiment to the multimode case. 
	The basic layout of this experiment underlies boson sampling---a promising model of nonuniversal quantum computations.
	Peculiarities of photocounting probabilities in such an experiment witness important nonclassical properties of electromagnetic field related to indistinguishability of boson particles.   
	In practice, these probabilities are changed from their theoretical values due to the imperfect ability of realistic detectors to distinguish numbers of bunched photons.
	We derive analytical expressions for photocounting distributions in the generalized Hong-Ou-Mandel experiment for the case of realistic photon-number resolving (PNR) detectors.
	It is shown that probabilities of properly postselected events are proportional to probabilities obtained for perfect PNR detectors. 
	Our results are illustrated with examples of arrays of on/off detectors and detectors affected by a finite dead time. 
\end{abstract}

\maketitle


\section{Introduction}
\label{Sec:Intro}


	Nonclassical phenomena in quantum optics are widely applied in both fundamental research and  modern quantum technologies.
	An important example is given by the Hong-Ou-Mandel effect \cite{hong1987}.  
	In the original configuration, it can be considered as an interference of two single photons on a beam splitter.
	If its transmittance and reflectance are equal, then photons cannot be registered at both outputs simultaneously---the effect  usually referred to as photon bunching. 
	Registration of bunched photons is possible for unbalanced beam splitters.
	However, the photocounting statistics in such cases still demonstrate strong peculiarities caused by the indistinguishability of boson particles.           

\begin{figure}[ht!]
			\includegraphics[width=1\linewidth]{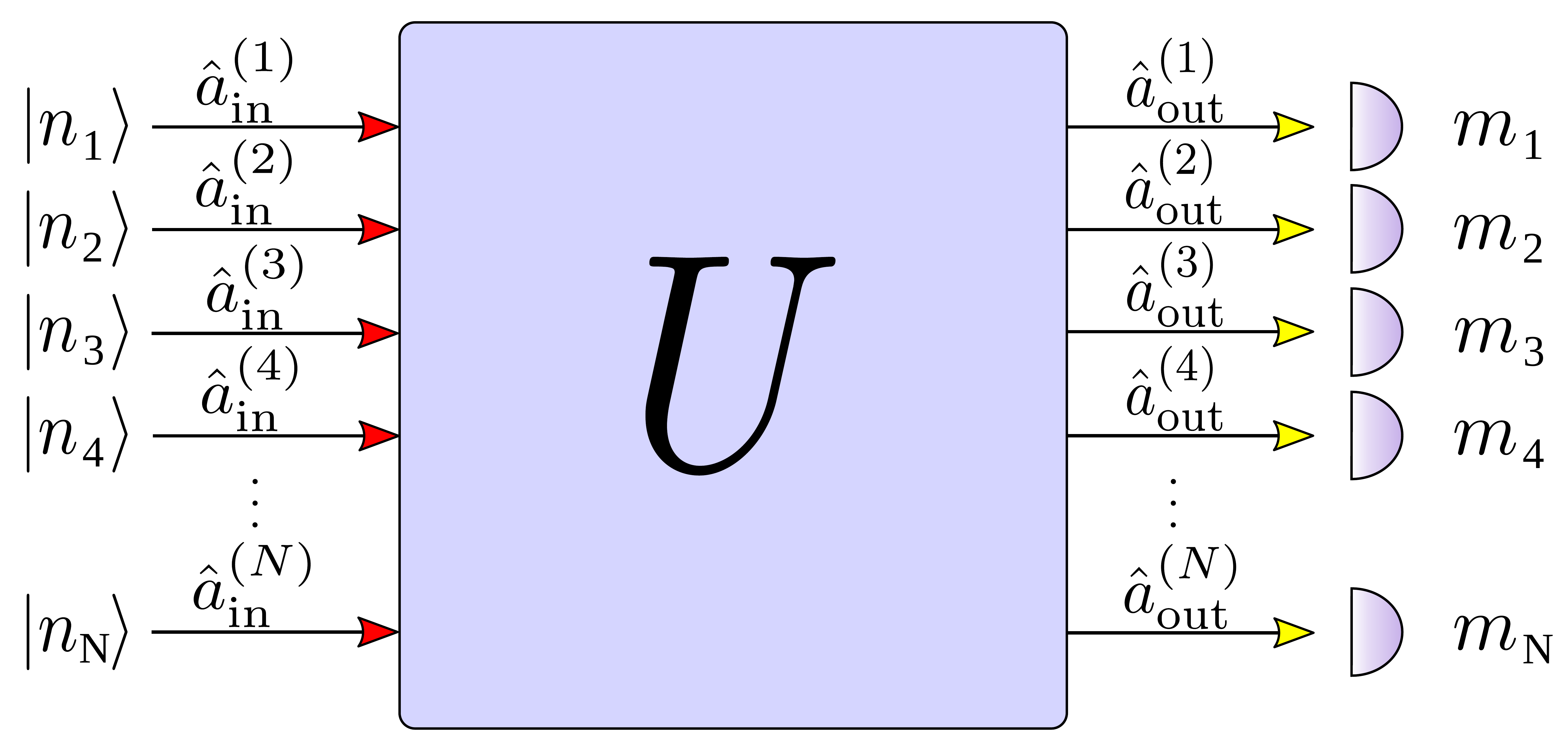}
			\caption{\label{Fig:BS_Scheme}
				The generalized Hong-Ou-Mandel experiment is depicted.
				$N$ field modes are coupled on a linear lossless interferometer described by the unitary matrix $U$.
				The $i$th mode is prepared in the Fock state $\ket{n_i}$.
				The output modes are detected by $N$ lossless PNR detectors.
			}
		\end{figure}

	An important generalization of the original Hong-Ou-Mandel experiment \cite{scheel2004,scheel2008,Lim2005} includes a multimode intereferometry, see figure~\ref{Fig:BS_Scheme}.  
	Let us consider $N$ input modes of optical radiation transformed by a linear interferometer as
		\begin{equation}\label{Eq:UnitaryTransform}
		\hat{a}_\mathrm{out}^{(j)}=\sum_{i=1}^{N}U_{ji}\hat{a}_\mathrm{in}^{(i)}. 
		\end{equation}
	Here $\hat{a}_\mathrm{in}^{(i)}$ and $\hat{a}_\mathrm{out}^{(j)}$ are field annihilation operators of input and output modes, respectively, and $U_{ji}$ is a unitary transformation matrix.
	Each input mode is prepared in the Fock state $\ket{n_i}$ such that the total number of input photons fulfills the condition
		\begin{equation}\label{Eq:PhotonPreserving}
		n=\sum_{i=1}^{N}n_i\leq N.  
		\end{equation}
	Each output mode is analyzed by a photon-number resolving (PNR) detector giving random outcomes $m_i$. 

	In the lossless scenario, the photon-number distribution at the output reads \cite{scheel2004,scheel2008}
		\begin{equation}\label{Eq:PND_Ideal}
		P_{m_1 \ldots m_N}=\frac{\big|\Perm U[1^{m_1} \ldots N^{m_N}|1^{n_1}\ldots N^{n_N}]\big|^2}{m_1!\ldots m_N!n_1!\ldots n_N!}.
		\end{equation}	
	Here $\Perm U[1^{m_1} \ldots N^{m_N}|1^{n_1}\ldots N^{n_N}]$ is the permanent of the matrix constructed from the elements of the original matrix $U_{ij}$ such that its row index $i$ and the column index $j$ appear $m_i$ and $n_j$ times, respectively.
	Calculation of permanents for matrices with complex entries is an example of \#P-hard problem.
	Hence, photocounting distributions (\ref{Eq:PND_Ideal}) cannot be efficiently calculated with classical devices.

	As it has been shown by Aaronson and Arkhipov \cite{aaronson2013}, the computational hardness is also related to classical sampling of events expressed in terms of the probabilities $P_{m_1 \ldots m_N}$ from eq.~(\ref{Eq:PND_Ideal}) for $n\ll N$.
	In particular, the computational hardness is considered for the collision-free regime, which takes place if $N=O\left(n^2\right)$, see reference~\cite{arkhipov2012}.
	Sampling such events within the generalized Hong-Ou-Mandel experiment may efficiently solve a computationally hard problem.
	This idea is the essence of boson sampling, which is an example of nonuniversal quantum computation.  


	Practical issues of the generalized Hong-Ou-Mandel experiment have been extensively discussed in literature due to its applications in boson-sampling schemes.
	These issues can be subdivided into three groups by relation to sources, interferometer, and detectors.
	Firstly, mode mismatch causes photons to be partially distinguishable, i.e. the interference between them is lost \cite{Rohde2012b,Tillmann2015,Renema2018,Moylett2018,Moylett2019,Shi2021}.    
	Secondly, unavoidable imperfections in the interferometer may affect the final results of sampling \cite{Leverrier2015}.
	Finally, the issues related to detector imperfections have been considered, such as detection losses \cite{aaronson2013,Rohde2012b}.   
	
	
	We address an issue related to detector imperfections but in a scenario where bunched photons play a crucial role in the generalized Hong-Ou-Mandel experiment.
	Indeed, photon bunching is an important aspect of the original Hong-Ou-Mandel experiment. 
	Its experimental observation plays a fundamental role in studying nonclassical properties of quantum light.
	It is also important that the true collision-free regime with $N>n^2$, cf. reference~\cite{arkhipov2012}, is hard to implement experimentally for large $n$.
	Therefore, these events will take place in realistic scenarios.
	According to reference~\cite{aaronson2013}, collision events can be useful for verification of eq.~(\ref{Eq:PND_Ideal}).
	Finally, we note a perspective for collision events to be applied to protocols of boson sampling validation since they enable us to get more information about the setup.
	

	A problem arising with practical consideration of collision events is that presently available detectors cannot distinguish perfectly between numbers of bunched photons.
	Consequently, eq.~(\ref{Eq:PND_Ideal}) does not hold anymore for realistic setups.   
	There exist several experimental techniques which enable to solve this problem at least approximately.
	The first method consists in separating the light beam into spatial \cite{paul1996,castelletto2007,schettini2007,blanchet08} or temporal \cite{achilles03,fitch03,rehacek03} modes and detecting each of them with an on/off detector. 
	The number of obtained clicks or triggered detectors can be approximately associated with the number of photons.
	The theory of such a detection has been presented in reference \cite{sperling12a}. 
	The other technique assumes counting the number of photocurrent pulses appearing inside a measurement time window. 
	Each pulse is interpreted as an evidence of a detected photon.
	However, this correspondence is an approximation since detectors cannot count any photons during their dead-time intervals after each pulse.
	Classical photocounting theory for this detection scheme has been presented earlier in references \cite{ricciardi66,muller73,muller74,cantor75,teich78,vannucci78,rapp2019}. 
	

	In this paper, we analyze the effect of realistic photon-number resolution on the phtocounting statistics in the generalized Hong-Ou-Mandel experiment.
	We demonstrate that a proper postselection can still be useful for sampling them from probabilities expressed via permanents appearing in eq. (\ref{Eq:PND_Ideal}).
	Our results are illustrated with arrays of on/off detectors and with single-photon detectors affected by dead time.


	The rest of the paper is organized as follows.
	In Sec.~\ref{Sec:IPNR} we present the general consideration of detectors with realistic photon-number resolution and discuss the corresponding photocounting statistics in the generalized Hong-Ou-Mandel experiment.
	In Sec.~\ref{Sec:ArrayDet} our results are applied to arrays of on/off detectors.
	The effect of detector dead time is considered in Sec.~\ref{Sec:DeadTime}.
	Summary and concluding remarks are given in Sec. \ref{Sec:Concl}.


\section{Realistic photon-number resolution}
\label{Sec:IPNR}

	In this section, we address general issues related to the generalized Hong-Ou-Mandel experiment with realistic photon-number resolution.
	Let us consider photon-number distribution at the output ports of the interferometer assuming the usage of ideal PNR detectors \cite{scheel2004,scheel2008},
		\begin{equation}\label{Eq:PhStat}
			P_{m_{1} \ldots m_{N}} = \Tr\big( \hat{\rho}\ket{m_1}\bra{m_1}\otimes\ldots\otimes\ket{m_N}\bra{m_N}\big).
		\end{equation}	
	Here $\hat{\rho}$ is the density operator of light modes at the interferometer outputs, and $\ket{m_i}\bra{m_i}$ are projectors on the Fock states.
	An important property of this distribution is that it has non-zero values only if the total number of detected photons, 
		\begin{equation}\label{Eq:PhotonPreserving2}
		n=\sum_{i=1}^{N}m_i,
		\end{equation}
	is exactly the same as the number of injected photons, cf. eq. (\ref{Eq:PhotonPreserving}).
	
	As the next step, we look into realistic photon-number resolution.
	In the most general case, the photocounting distribution is given by
		\begin{equation}\label{Eq:PNS}
			\rho_{k_{1} \ldots k_{N}} = \Tr\left( \hat{\rho}\,\hat{\Pi}_{k_1}\otimes\ldots\otimes\hat{\Pi}_{k_N}\right).
		\end{equation}
	Here $\hat{\Pi}_{k}$ is the positive operator-valued measure (POVM) \cite{nielsen10} describing the photocounting procedure.  
	This POVM can always be expanded by projectors on the Fock states \cite{kovalenko2018} as
		\begin{equation}\label{Eq:ClickVsPhotonPOVMs}
			\hat{\Pi}_{k}=\sum_{m=0}^{+\infty}P_{k|m}\ket{m}\bra{m},
		\end{equation}	
	where the coefficients 
		\begin{equation}\label{Eq:ConP}
			P_{k|m}=\bra{m}\hat{\Pi}_{k}\ket{m}
		\end{equation}
	can be interpreted as the probabilities to get $k$ counts, e.g. clicks or pulses, given $m$ injected photons. 
	In the following sections we will obtain these coefficients for specific photocounting techniques.
	In practice, they can also be reconstructed from the measurement data given certified Fock-state sources. 
	Substitution of eq.~(\ref{Eq:ClickVsPhotonPOVMs}) into eq.~(\ref{Eq:PNS}) yields
		\begin{equation}\label{Eq:RealStatGen}
			\rho_{k_{1} \ldots k_{N}} =\sum_{\sum\limits_{i=1}^{N} m_i=n}P_{k_1|m_1}\ldots P_{k_N|m_N}P_{m_{1} \ldots m_{N}},
		\end{equation}
	where sum is taken over the indices $m_i$ obeying the condition (\ref{Eq:PhotonPreserving2}).
	It is worth noting that summation over $\{m_i\}$ satisfying the given condition holds only for Fock states at the input.
	In a more general scenario of arbitrary states, this summation should be taken for all possible values of $\{m_i\}$.
  	
  	The situation is drastically simplified if we assume that all counts of the considered detectors are solely related to absorbed signal photons, i.e. no dark counts, afterpulses, etc. take place.
  	This yields $P_{k|m}=0$ for $m<k$, which means that the number of photocounts does not exceed the number of photons.
  	We restrict our consideration by the measurement outcomes with the total number of photocounts being equal to the number of injected photons, i.e. condition (\ref{Eq:PhotonPreserving2}) with $m_i$ replaced by $k_i$ is satisfied.
  	In this case, the terms with at least one $P_{k_i|m_i}$ such that $k_i{<}m_i$ in  eq.~(\ref{Eq:RealStatGen}) necessarily include another constituent $P_{k_j|m_j}$ with $k_j{>}m_j$, which is equal zero.
  	This implies that eq.~(\ref{Eq:RealStatGen}) contains only one nonzero term, i.e. this equation is reduced to the form
  		\begin{equation}\label{Eq:RealStat}
			\rho_{k_{1} \ldots k_{N}}= C_{k_{1} \ldots k_{N}}P_{k_1 \ldots k_N},
		\end{equation}
	where
		\begin{equation}\label{Eq:CorrectCoeff}
			C_{k_{1} \ldots k_{N}}=\prod\limits_{i=1}^{N}P_{k_i|k_i}	
		\end{equation}
	are the correction coefficients between photocounting statistics of ideal and realistic detectors.
	Therefore, eq.~(\ref{Eq:RealStat}) directly links sampling probabilities of realistic and ideal PNR detectors.
	The latter are expressed in terms of permanents appearing in eq.~(\ref{Eq:PND_Ideal}).
	The total efficiency of the corresponding postselection, i.e. the probability that the number of counts is equal to the number of injected photons, is given by sum of all $\rho_{k_{1} \ldots k_{N}}$ satisfying condition (\ref{Eq:PhotonPreserving2}).
	In the general case, its calculation involves values of permanents and thus cannot be evaluated with classical devices. 
		
	An elementary example is given by losses in the PNR detectors.
	The corresponding POVM reads
		\begin{equation}
			\hat{\Pi}_{k}=\hat{F}_{k}(\eta)=
			\normord{\frac{\left(\eta\hat{n}\right)^{k}}{k!}\exp\left(-\eta\hat{n}\right)},
			\label{Eq:POVM_losses}
		\end{equation}
	where $\eta\in[0,1]$ is the detection efficiency, $\hat{n}$ is the photon-number operator, and $:\ldots:$ means normal ordering.
	In this case, $P_{k_i|k_i}=\eta^{k_i}$, see references \cite{Scully1969,Lee2005,Semenov2008}.
	Hence, each correction coefficient is given by
		\begin{equation}\label{Eq:CorrLosses}
			C_{k_{1} \ldots k_{N}}=\eta^n,
		\end{equation}
	which coincides with the total postselection efficiency.
	A distinctive feature of this example is that the correction coefficients are the same for all events with $k_i$ obeying eq.~(\ref{Eq:PhotonPreserving2}).
	Therefore, the photon-number statistics~(\ref{Eq:PND_Ideal}) is proportional to the statistics of properly postselected events.
	
	This property does not hold for the general realistic PNR detectors.
	According to eq.~(\ref{Eq:RealStat}), each probability $P_{k_1 \ldots k_N}$ should be multiplied with the correction coefficient $C_{k_{1} \ldots k_{N}}$ corresponding to the special values of $k_{1}, \ldots, k_{N}$.
	This implies that eq.~(\ref{Eq:RealStat}) can be used for reconstruction of probabilities $P_{k_1 \ldots k_N}$ and thus for verification of eq.~(\ref{Eq:PND_Ideal}) in the case of appropriate photon numbers as it is discussed in reference~\cite{aaronson2013}. 
	
	The values of the correction coefficients $C_{k_{1} \ldots k_{N}}$ for the general realistic PNR detectors depend only on $k_i$ different from $0$ and $1$.
	This is based on the fact that $P_{0|0}=1$ and $P_{1|1}=\eta$.
	Consider two correction coefficients $C_{k_{1} \ldots k_{N}}$ and $C_{l_{1} \ldots l_{M}}$.
	In general, $N$ may not be equal to $M$.
	Let for a number $Q>0$  there exist such sets $\{i_q|q=1\ldots Q\}$ and $\{j_q|q=1\ldots Q\}$ that $k_{i_q}=l_{j_q}$.
	All others $k_i$ and $l_j$ are equal to 0 or 1.
	In this case, the coefficients are equal, and we can use for them a single identifier
		\begin{equation}\label{Eq:CorCeff_Simpl}
			C_{\{k_{q_1}\ldots k_{q_Q}\}}\equiv C_{k_{1} \ldots k_{N}}=C_{l_{1} \ldots l_{M}},
		\end{equation}
	depending only on the set of ${\{k_{q_1}\ldots k_{q_Q}\}}$ with elements differ from 0 and 1.


\section{Arrays of on/off detectors}
	\label{Sec:ArrayDet}
	
\begin{figure}[h]
    		\includegraphics[width=1\linewidth]{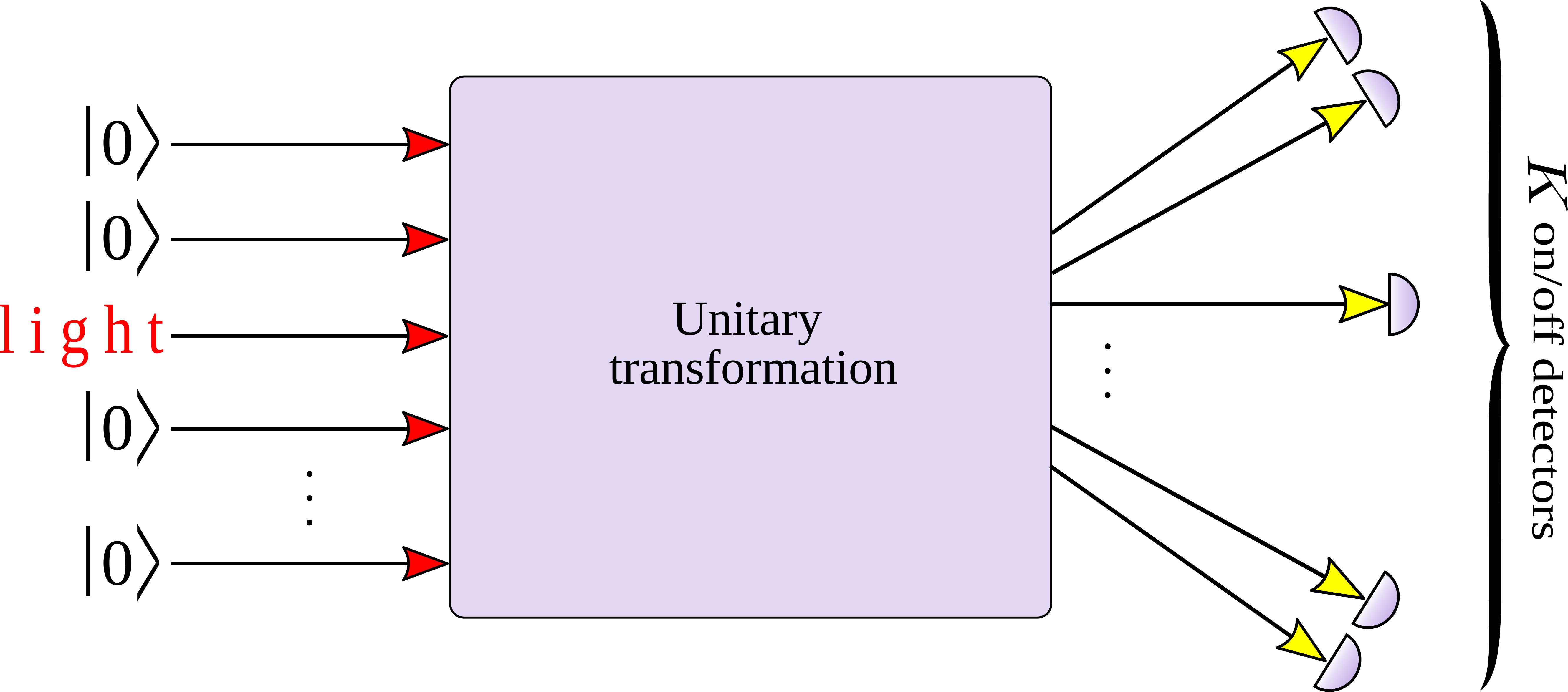}
    		\caption{\label{Fig:ArrayDet}
    			An array of on/off detectors is depicted.
    			The light beam is split into $K$ equal beams.
    			Each beam is detected by an on/off detector.
    			The measurement outcomes are given by the numbers of triggered detectors.
    }
    	\end{figure}

	As it has been already discussed in Introduction, arrays of on/off detectors \cite{paul1996,castelletto2007,schettini2007,blanchet08} represent an efficient experimental tool enabling an approximate resolution between numbers of photons.
	The corresponding measurement layout, see figure~\ref{Fig:ArrayDet}, is based on splitting  a light beam into $K$ spatial modes and detecting each of them by an on/off detector.
	For such a scenario, the measurement outcome is given by the number of triggered detectors.
	Equivalently, one can use a temporal separation of light modes by loops of optical fibers as it is discussed in references~\cite{achilles03,fitch03,rehacek03}.
	In this Section, we consider measurement outcomes of the generalized Hong-Ou-Mandel experiment, assuming that each output mode of the interferometer is analyzed by an array of on/off detectors or by an equal scheme with time-mode separation.   
	
	The POVM for this detection scheme \cite{sperling12a} is given by
		\begin{equation}
        	\hat{\Pi}_{k}={K\choose k}\normord{\left(1-e^{-\eta\hat{n}/K}\right)^{k}
        		e^{-\eta\hat{n}(K-k)/K}}.
        	\label{Eq:POVM_Array}
    	\end{equation}
	Here $K$ is the number of on/off detectors in the array, $k=0\ldots K$ is the number of triggered detectors, and $\eta$ is the detection efficiency.
   	In this case, the probability to get $k_i$ clicks given $k_i$ photons, cf. eq. (\ref{Eq:ConP}), is directly obtained \cite{sperling12a,kovalenko2018} as
    	\begin{equation}
            P_{k_i| k_i}=\left(\frac{\eta}{K}\right)^{k_i}\frac{K!}{(K-k_i)!}. 
    	\end{equation}
 	Substituting it into eq. (\ref{Eq:CorrectCoeff}) we arrive at the corresponding correction coefficients,
    	\begin{equation}\label{Eq:CorCoeff_Array}
        	C_{k_{1} \ldots k_{N}}=\left(\frac{\eta}{K}\right)^n\prod\limits_{i=1}^{N}\frac{K!}{(K-k_i)!}.
    	\end{equation}
    In the form of eq.~(\ref{Eq:CorCeff_Simpl}) it can be rewritten as
        \begin{equation}\label{Eq:CorCoeff_Array1}
    		C_{\{k_{q_1}\ldots k_{q_Q}\}}=\eta^n\prod\limits_{i=1}^{Q}\frac{K!}{K^{q_i}(K-k_{q_i})!}.
    	\end{equation}
	This expression includes the multiplier related to the detector losses given by eq.~(\ref{Eq:CorrLosses}).
	In addition, it explicitly depends on the set of numbers $k_i$, which is not the case for ordinary losses. 
	
\begin{figure}[h]
        	\includegraphics[width=1\linewidth]{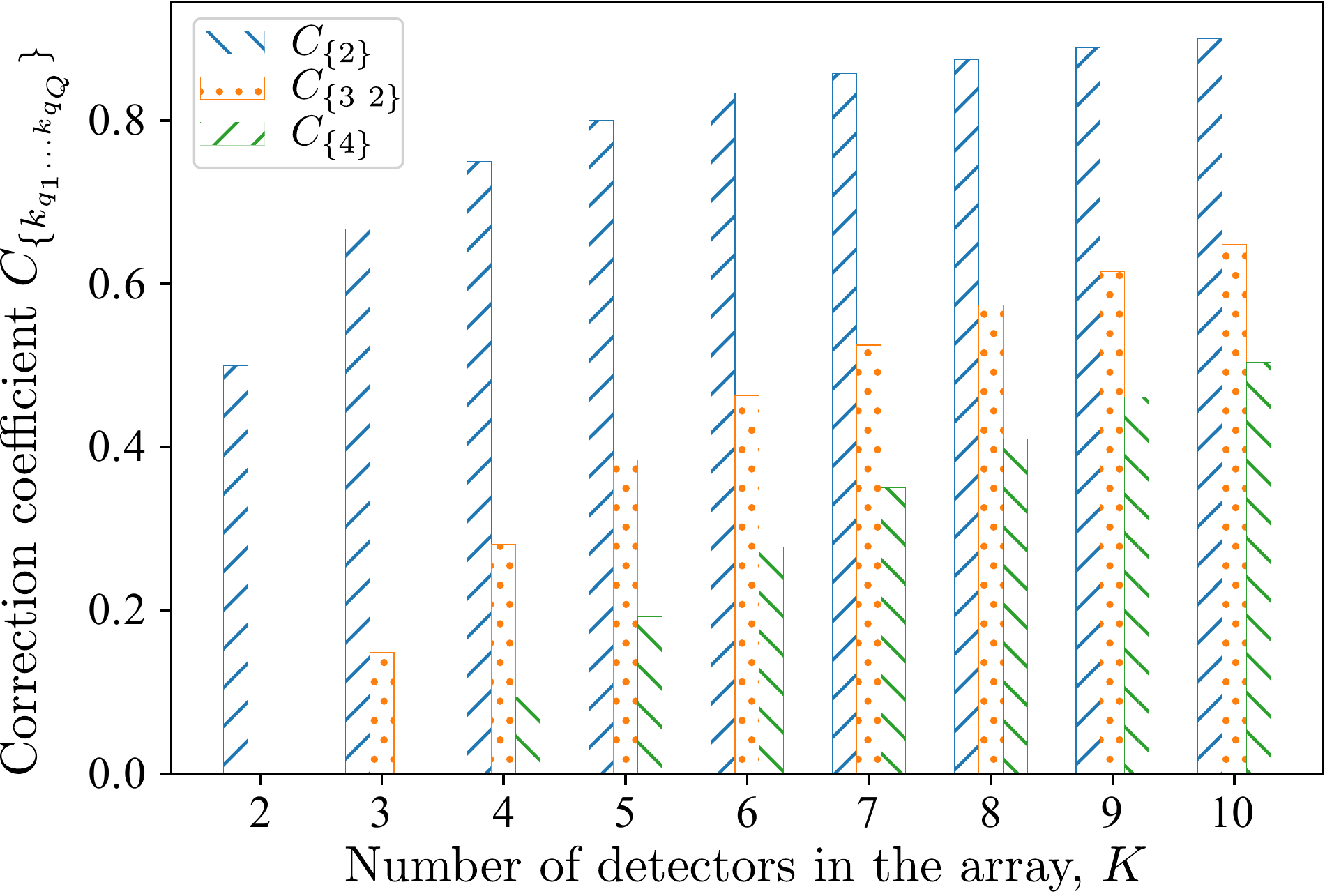}
        	\caption{\label{Fig:diagram_array} The diagram shows the values of the correction coefficients (\ref{Eq:CorCoeff_Array1}) for a six-port interferometer depending on the number $K$ of on/off detectors in the array.
        	Without loss of generality, we assume $\eta=1$.}

        \end{figure}

	The correction coefficients (\ref{Eq:CorCoeff_Array}) characterize the detector impact on the photocounting statistics at the interferometer output.
	It establishes a link between data obtained with arrays of on/off detectors and photon-number statistics related to ideal PNR detectors. 
	The dependence of the correction coefficients (\ref{Eq:CorCoeff_Array}) on the number $K$ of on/off detectors in the array is shown in figure~\ref{Fig:diagram_array}.
	Here we assume $\eta=1$ without loss of generality.
	This coefficient is equal to unity if $k_i=0,1$ for all $i=1\ldots N$, which corresponds to the standard boson-sampling scheme in the collision-free regime.
	In all other cases, the correction coefficient has non-trivial dependence on the set of $k_i$.
	The general rule is that these coefficients rapidly vanish while the maximal value of $k_i$ in the set increases.
	When growing the number $K$ of on/off detectors in the array, the correction coefficients approach unity.
	However, in the most cases its contribution must be taken into account.
	
\section{Detection with finite dead time}
\label{Sec:DeadTime}
  
    In this section we consider the generalized Hong-Ou-Mandel experiment with a method based on counting photons detected during a measurement time window $\tau_\mathrm{m}$.
    This time characterizes a nonmonochromatic mode of light, the quantum state of which is analyzed. 
    The light incident on a single-photon detector induces photocurrent pulses.
    Each pulse can be related to a single detected photon.
    The measurement outcome of such a scheme is the number of pulses appearing during the measurement time window $\tau_\mathrm{m}$.  
    
\begin{figure}[h]
	    	\includegraphics[width=1\linewidth]{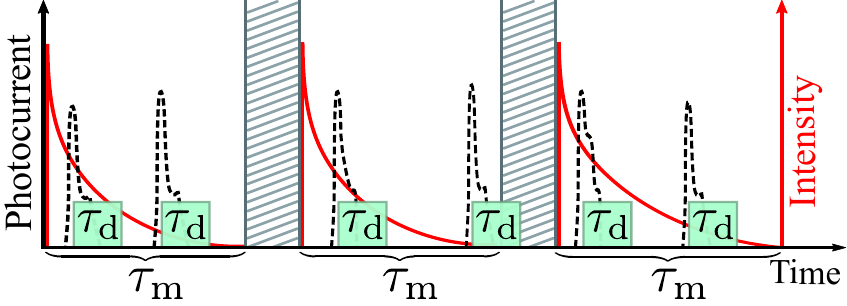}
	    	\caption{\label{Fig:FinDT}
	    		The idea of measurements with independent time windows is depicted.
	    		A light mode (solid line) arrives at the beginning of the time-window $\tau_\mathrm{m}$.
	    		Detected photons induce photocurrent pulses (dashed line).
	    		After each pulse, detector cannot register photons during the dead time $\tau_\mathrm{d}$ (shaded area).
	    		Each time window is followed by a time interval (hatched area) of darkened detector input exceeding $\tau_\mathrm{d}$.}
	    \end{figure}

    The main problem of the considered detection type is that the photons cannot be detected during a dead-time interval $\tau_\mathrm{d}$ that follows each pulse. 
    This results in non-ideal photon-number resolution.
    Another problem appears when the dead-time interval related to the last pulse overlaps with the next measurement time window.
    This leads to a nonlinear memory effect in photocounting statistics. 
    In order to avoid it and make time windows independent, detector input should be darkened after each measurement time window, see figure~\ref{Fig:FinDT}.  
    
    In the context of the generalized Hong-Ou-Mandel experiment, we address two scenarios related to this type of detection.
    First, we consider the case when the input light is a continuous monochromatic wave.
    It is split into parts, each of them related to a measurement time window.
    Next, we consider a more realistic scenario in which Fock-state sources irradiate nonmonochromatic modes of light.
    As an example, we consider light modes in the form of exponential decay.

\subsection{POVM for the detection with dead time}    
\label{Sec:POVM}

	In this subsection, we derive the POVM for detection with independent time windows.
	In the most general scenario, the light mode inside the measurement time interval $[0,\tau_\mathrm{m}]$ is nonmonochromatic; see, e.g., reference~\cite{mandel_book} for analysis of nonmonochromatic modes.
	The field intensity is a function of time, $I(t)$, normalized in the measurement time window by the detection efficiency $\eta$, 
		\begin{equation}\label{Eq:NormInt}
		\int_0^{\tau_\mathrm{m}} I(t)dt=\eta.
		\end{equation}
	The POVM should explicitly include this function.
	It can be considered as a particular case of the POVM derived in reference~\cite{Uzunova2022}.
	For the sake of clarity and completeness, we adapt this  derivation procedure  to the considered detection scenario.  

	Instead of the operator form of the POVM, $\hat{\Pi}_{k}$,  we can use its phase-space $Q$ symbol defined as the average with coherent states $\ket{\alpha}$, cf. references~\cite{cahill69,cahill69a},
		\begin{equation}
			\Pi_{k}(\alpha)=\bra{\alpha}\hat{\Pi}_{k}\ket{\alpha}.	
		\end{equation}	
	The original operator form can be reconstructed from the $Q$ symbol using the rule
		\begin{equation}\label{Eq:Rule}
			\bra{\alpha}:\hat{f}\left(\hat{a},\hat{a}^\dag\right):\ket{\alpha}=f\left(\alpha,\alpha^\ast\right),
		\end{equation}
	where $\hat{a}$ and $\hat{a}^\dag$ are field annihilation and creation operators, respectively.
	The $Q$ symbols of the POVM can be interpreted as photocounting distributions for the coherent state $\ket{\alpha}$.
	
	For $k=0$ the POVM element is given by $\hat{\Pi}_0=\hat{F}_0(\eta)$, cf. eq.~(\ref{Eq:POVM_losses}).  
	The $Q$ symbols of the POVM for $k\geq1$ can be obtained via integration of the unnormalized probability density $\pi_{k}\left(\mathbf{t}|\alpha\right)$ to register $k$ pulses in the time moments $\mathbf{t}=(t_1,\ldots,t_{k})$ given the coherent state $\ket{\alpha}$,  
		\begin{equation}\label{Eq:POVM_Q_gen}
			\Pi_{k}(\alpha)=\int_{T_{k}}\D^{k} \mathbf{t}\, \pi_{k}\left(\mathbf{t}|\alpha\right).
		\end{equation} 
	where $\mathbf{t}\in T_{k}$ satisfies the condition $0\leq t_1\leq t_2\ldots\leq t_{k}\leq \tau_\mathrm{m}$.
	In order to derive the unnormalized probability density $\pi_{k}\left(\mathbf{t}|\alpha\right)$ we start with the example of $\pi_1\left(t_1|\alpha\right)$. 
	For this purpose, we consider the infinitesimal probability to get a pulse at the time $t_1$ during the time-interval $\D t_1$,
		\begin{equation}
			|\alpha|^2I(t_1)\D t_1\exp\left(-|\alpha|^2I(t_1)\D t_1\right)\rightarrow |\alpha|^2I(t_1)\D t_1.
		\end{equation}
	It should be multiplied with the probabilities to get no pulses before and after the time moment $t_1$,
		\begin{equation}
			\exp\left(-|\alpha|^2\int\limits_{0}^{t_1}\D t I(t)\right)
		\end{equation}
	and
		\begin{equation}
			\exp\left(-|\alpha|^2\int\limits_{t_1}^{\tau_\mathrm{d}}\D t I(t)\theta(t-t_1-\tau_\mathrm{d})\right),
		\end{equation}	
	respectively.
	Here $\theta(t-t_1-\tau_\mathrm{d})$ is the Heaviside step function, which describes the impossibility of appearing pulses during the dead-time interval $\tau_\mathrm{d}$ after the time moment $t_1$.
	Introducing the function 
		\begin{equation}\label{Eq:Xi_1}
			\Xi_1\left(t_1\right)=\int\limits_{0}^{t_1}\D t I(t)
			+\int\limits_{t_1}^{\tau_\mathrm{m}}\D t I(t)\theta(t-t_1-\tau_\mathrm{d}),
		\end{equation}	
	we derive the expression
		\begin{equation}\label{Eq:POBM1}
			\pi_1\left(t_1|\alpha\right)={|\alpha|^{2}} I\left(t_1\right)\exp\left[-|\alpha|^2\Xi_1\left(\mathbf{t}\right)\right]
		\end{equation}  	
	for the probability distribution to get a pulse at the time moment $t_1$ given the coherent state $\ket{\alpha}$.
	
	The unnormalized probability density $\pi_{k}\left(\mathbf{t}|\alpha\right)$ to register $k\geq2$ pulses in the time moments $\mathbf{t}=(t_1,\ldots,t_{k})$ given the coherent state $\ket{\alpha}$ is constructed in a similar way. It is composed of the following constituents:

		\begin{itemize}
			\item The unnormalized probability density of registering a pulse in the time moment $t_1$ given by $|\alpha|^2I(t_1)$;
			\item The unnormalized probability density of registering pulses at the time moments $t_i$ for $i=2\ldots {k}$  given by $|\alpha|^2I(t_i)$;
			\item The zero-probability for the $i$th pulses to appear after the $(i-1)$th pulses during the dead time $\tau_\mathrm{d}$ described by the Heaviside step function $\theta(t_i-t_{i-1}-\tau_\mathrm{d})$; 
			\item The probability of registering no photons from the time moment $t=0$ and up to the first pulse in the time moment $t_1$ given by $\exp\left(-|\alpha|^2\int_{0}^{t_1}\D tI(t)\right)$;
			\item The probability of registering no photons in the time domain between the $i$th and $(i+1)th$ pulses given by $\exp\left(-|\alpha|^2\int_{t_i}^{t_{i+1}}\D tI(t)\theta(t-t_i-\tau_\mathrm{d})\right)$, where the dead time is taken into account;
			\item The probability of registering no photons in the time domains between the ${k}$th pulse and the time moment $t=\tau_\mathrm{m}$ given by $\exp\left(-|\alpha|^2\int_{t_{k}}^{\tau_\mathrm{m}}\D tI(t)\theta(t-t_{k}-\tau_\mathrm{d})\right)$, where the dead time is taken into account.
		\end{itemize}
	Combining all these elements we obtain the unnormalized probability density as
		\begin{equation}\label{Eq:POVM_DT_gen}
			\pi_{k}\left(\mathbf{t}|\alpha\right)={|\alpha|^{2{k}}} \mathcal{I}_{k}\left(\mathbf{t}\right)\exp\left[-|\alpha|^2\Xi_{k}\left(\mathbf{t}\right)\right],
		\end{equation}  
	where
		\begin{equation}\label{Eq:I}
			\mathcal{I}_{k}\left(\mathbf{t}\right)=I(t_1)\prod\limits_{i=2}^{{k}}I(t_i)\theta\left(t_i-t_{i-1}-\tau_\mathrm{d}\right),
		\end{equation}
	and
		\begin{eqnarray}
			\Xi_{k}\left(\mathbf{t}\right)=\int\limits_{0}^{t_1}\D t I(t)&+\sum\limits_{i=1}^{{k}-1}\int\limits_{t_i}^{t_{i+1}}\D t I(t)\theta(t-t_i-\tau_\mathrm{d})\nonumber\\
			&+\int\limits_{t_{k}}^{\tau_\mathrm{m}}\D t I(t)\theta(t-t_{k}-\tau_\mathrm{d}).\label{Eq:Xi}
		\end{eqnarray}
	Equation (\ref{Eq:POVM_DT_gen}) can also be applied for ${k}=1$, where $\Xi_1\left(t_1\right)$ is given by eqs.~(\ref{Eq:Xi_1}) and $\mathcal{I}(t_1)=I(t_1)$.
		
	The $Q$ symbols of the POVM are obtained via integration as it is given by eq.~(\ref{Eq:POVM_Q_gen}). 
	Applying the rule (\ref{Eq:Rule}), we obtain the POVM 
		\begin{equation}\label{POVM_NM}
			\hat{\Pi}_{k}=:{\hat{n}^{{k}}}\int_{T_{k}}\D^{k}\mathbf{t}\, \mathcal{I}_{k}\left(\mathbf{t}\right)\exp\left[-\hat{n}\Xi_{k}\left(\mathbf{t}\right)\right] :
		\end{equation}
	for ${k}=1\ldots K+1$.
	Here $K=[\tau_\mathrm{m}/\tau_\mathrm{d}]$ is the number of whole dead time intervals fitting inside the measurement time window.

\subsection{Monochromatic light}

   Let us consider a monochromatic light mode.
   In this case, the intensity is given by
   		\begin{equation}
   			I(t)=\frac{\eta}{\tau_\mathrm{m}}.
   		\end{equation}
   With this form of the intensity, eq.~(\ref{Eq:Xi}) can be integrated explicitly.
   Substituting the result of integration and eq.~(\ref{Eq:I}) into eq.~(\ref{POVM_NM}), we arrive at the POVM 
        \begin{equation}\label{Eq:POVM_DT_0}
        	\hat{\Pi}_{0}=\hat{F}_{0}\left(\eta\right),
        \end{equation}
        \begin{equation}\label{Eq:POVM_DT}
            \hat{\Pi}_{{k}}=\sum_{l=0}^{k} \hat{F}_{l}\left(\eta\eta_{k}\right)-\sum_{l=0}^{k-1} \hat{F}_{l}\left(\eta\eta_{k-1}\right)
        \end{equation}
    for $k=1\ldots K$,
         \begin{equation}\label{Eq:POVM_DT_NP1}
    		\hat{\Pi}_{K+1}=1-\sum_{l=0}^{K} \hat{F}_{l}\left(\eta\eta_{K}\right).
    	\end{equation}	
    Here $\hat{F}_{l}\left(\eta\right)$ is the POVM of the ideal PNR detectors with losses, cf. eq. (\ref{Eq:POVM_losses}), and
        \begin{equation}\label{Eq:AdjEff}
            \eta_{k}=\frac{\tau_\mathrm{m}-{k}\tau_\mathrm{d}}{\tau_\mathrm{m}}
    \end{equation}
    is the adjusting efficiency.
    These equations can be considered as a straightforward generalization of the corresponding results in the classical photodetection theory \cite{ricciardi66,muller73,muller74,cantor75,teich78,vannucci78,rapp2019}.
    
	In the considered scenario, the probability to get $k_i$ clicks given $k_i$ photons is obtained by substituting eqs.~(\ref{Eq:POVM_DT_0}), (\ref{Eq:POVM_DT}), and (\ref{Eq:POVM_DT_NP1}) into eq.~(\ref{Eq:ConP}),
    	\begin{equation}\label{Eq:CondProb1}
	        P_{k_i|k_i}=(\eta \eta_{k_i-1})^{k_i}.
	    \end{equation}
	Thus, the corresponding correction coefficients are given by
	    \begin{equation}
        	C_{k_1 \ldots k_N}=\eta^n \prod\limits_{i=1}^{N}\eta_{k_i-1}^{k_i}.
	        \label{Eq:CorCoeff_DT}
    	\end{equation}
    In the form of eq.~(\ref{Eq:CorCeff_Simpl}) it can be rewritten as
    	\begin{equation}
    		C_{\{k_{q_1}\ldots k_{q_Q}\}}=\eta^n \prod\limits_{i=1}^{Q}\eta_{k_{q_i}-1}^{k_{q_i}}.
    		\label{Eq:CorCoeff_DT1}
    	\end{equation}
	Similar to the case of array detectors, this expression includes the multiplier related to the detection efficiency, cf. eq. (\ref{Eq:CorrLosses}).
	At the same time, the correction coefficients are characterized by a non-trivial dependence on the set of numbers $k_i$.	
	
\begin{figure}[h!]
	    	\includegraphics[width=1\linewidth]{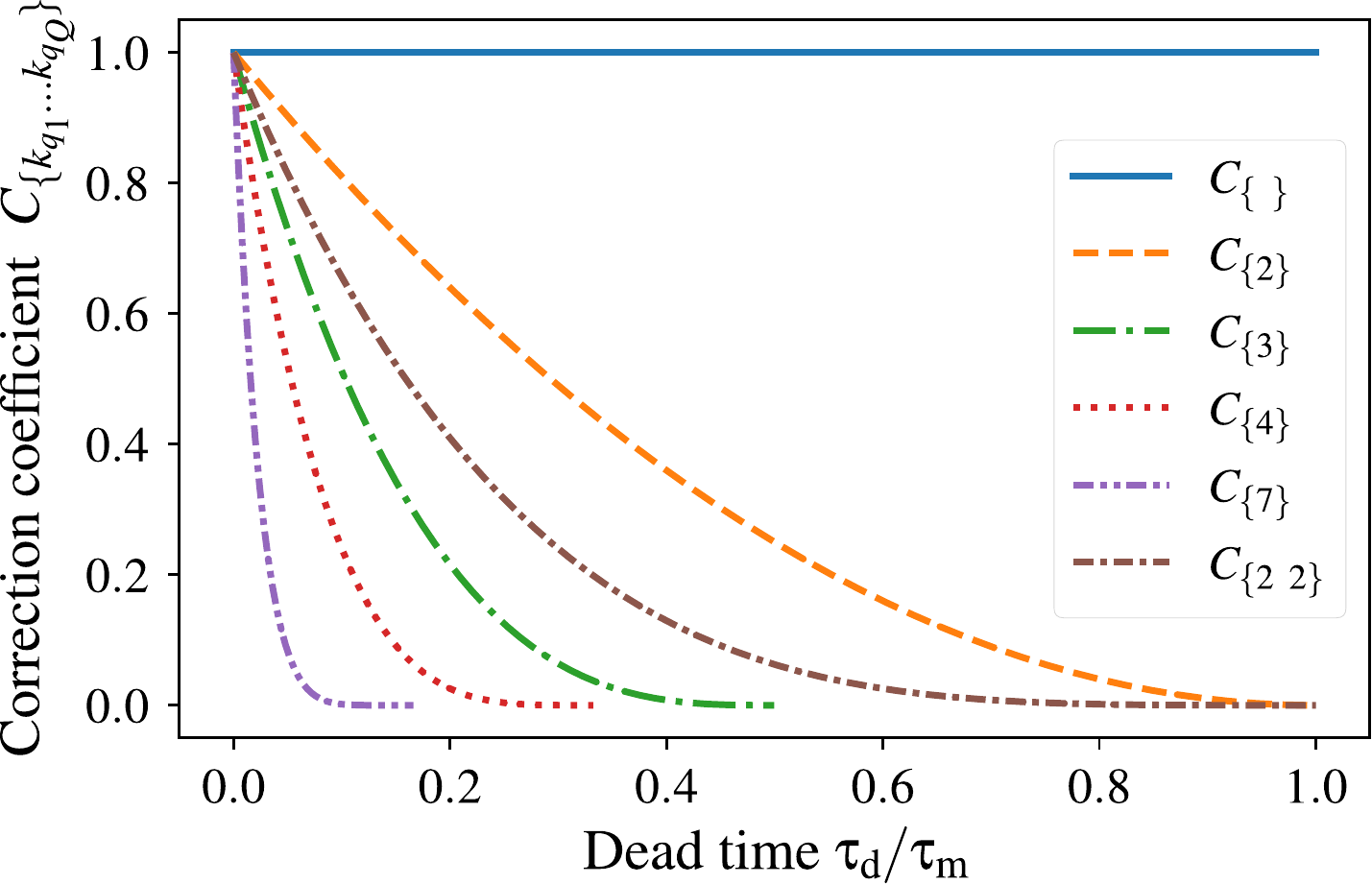}
	    	\caption{\label{Fig:GraphTD1}
	    	Dependence of the correction coefficients $ C_{\{k_{q_1}\ldots k_{q_Q}\}} $ on the dead time $ \tau_ \mathrm{d} $ in the scenario of monochromatic incident light.
    		Without loss of generality we assume $\eta=1$.
    		$C_{\{\hspace{0.3em}\}}$ corresponds to the collision-free events.
    		}
	    \end{figure}

	The dependence of the correction coefficients on the dead time is shown in figure~\ref{Fig:GraphTD1}, where we also assume $\eta=1$.
	It can be seen that the correction coefficient is equal to unity in two cases: when the dead time tends to zero and for the no-collision events, i.e. $k_i=0,1$ for $i=1\ldots N$.
	The correction coefficients vanish when the dead time and the maximal number $n_i$ increase.
	Moreover, the larger the maximal number in the set $\{k_{q_1}\ldots k_{q_Q}\}$, the faster the correction coefficient $ C_ {\{k_{q_1}\ldots k_{q_Q}\}} $ as a function of the dead time tends to zero.
	When the dead time $\tau_\mathrm{d}$ approaches the measurement time $\tau_\mathrm{m}$, all coefficients except for those with $k_i=0,1$ tend to zero.
	For small values of the correction coefficients, the sampling can be problematic since such events are rare.
	
\subsection{Nonmonochromatic modes of light}
\label{Sec:NML}
	
	Let us consider a scenario with identical nonmonochromatic field modes at the interferometer inputs.
	For example, it can be chosen in the form of exponential decay,
		\begin{equation}\label{Eq:ExpDecay}
			I(t)=\frac{\eta\gamma }{\tau_\mathrm{m}\left[1-\exp\left(-\gamma\right)\right]}\exp\left({-\gamma\frac{ t}{\tau_\mathrm{m}}}\right),
		\end{equation}	 
	where $\gamma>0$ is the decay rate.
	Such a mode can represent a pulse escaping from a leaky cavity, see e.g. \cite{lang1973,ujihara_book,Khanbekyan2004}.
	For $\gamma\rightarrow 0$, one gets $I(t)=1/\tau_\mathrm{m}$ that corresponds to the monochromatic field.
	Here we consider a simple scenario for which all nonmonochromatic modes at the interferometer inputs are identical and, consequently, can be considered as interfered.
	In a different case, when the modes are non-identical, the photons are partially distinguishable and interference is lost; see references~\cite{Rohde2012b,Tillmann2015,Renema2018,Moylett2018,Moylett2019,Shi2021}.

	Substitution of eq.~(\ref{POVM_NM}) into eq.~(\ref{Eq:ConP}) for $k=m=k_i$ yields the probability to get $k_i$ pulses given $k_i$ photons,
		\begin{equation}
			P_{k_i|k_i}=k_i!\int_{T_{k_i}}\D^{k_i}\mathbf{t}\, \mathcal{I}_{k_i}\left(\mathbf{t}\right).
		\end{equation}
	Applying the explicit form (\ref{Eq:ExpDecay}) of $I(t)$, we arrive at the analytical form of this probability,
		\begin{equation}
			P_{k_i|k_i}=\left[\eta\frac{\sinh\left(\gamma\eta_{k_i-1}/2\right)}{\sinh\left(\gamma/2\right)}\right]^{k_i},
		\end{equation}
	where $\eta_{k}$ is given by eq.~(\ref{Eq:AdjEff}).	
	For $\gamma\rightarrow 0$ this expression tends to eq.~(\ref{Eq:CondProb1}).

\begin{figure}[h!]
	    	\includegraphics[width=1\linewidth]{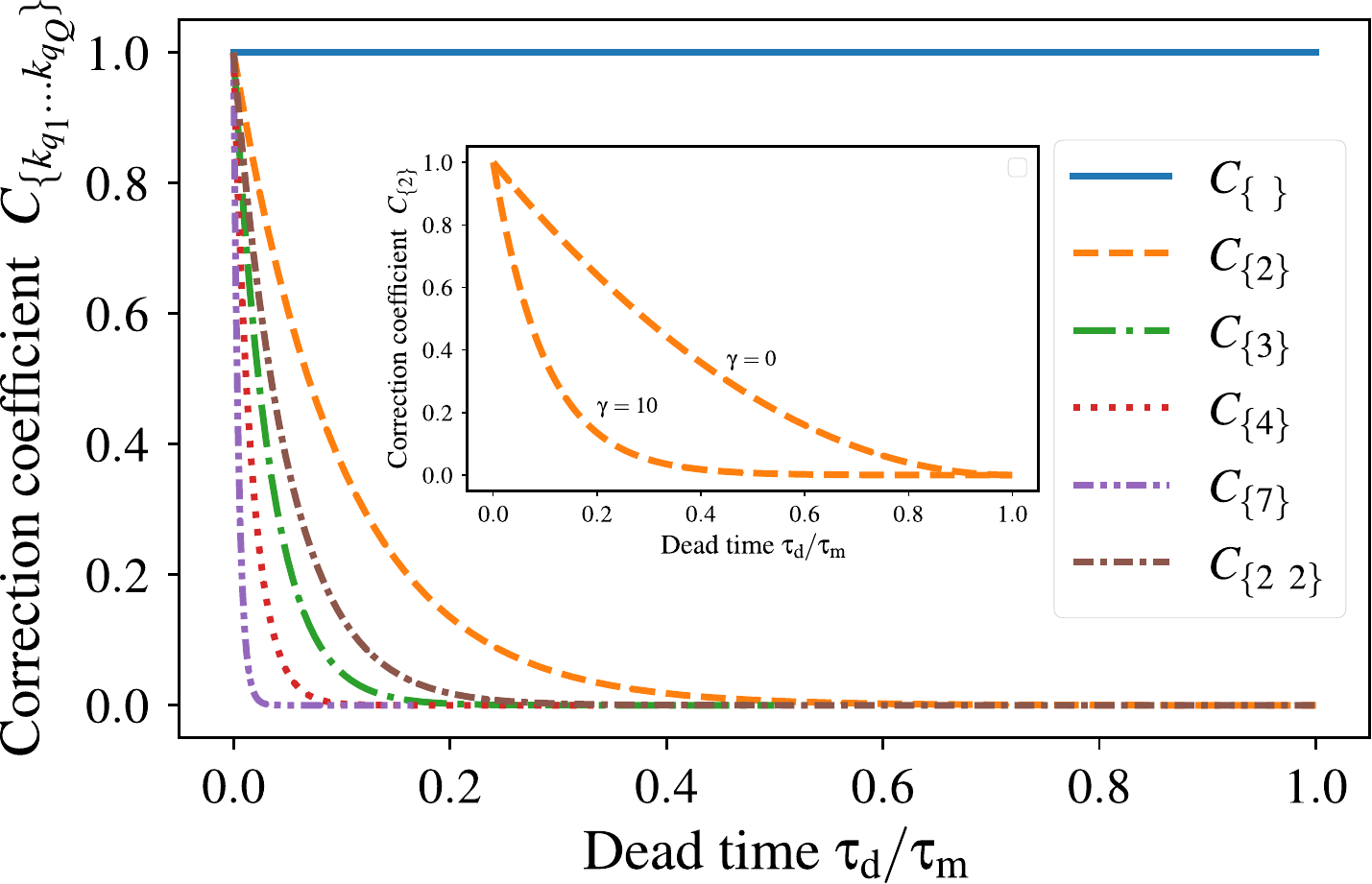}
	    	\caption{\label{Fig:GraphTD2}
	    	The same as in figure~\ref{Fig:GraphTD1} but for the scenario of nonmonochromatic mode in the form of eq.~(\ref{Eq:ExpDecay}) with  $\gamma=10$.
    		The inset shows the correction coefficient $C_{\{2\}}$ for the monochromatic ($\gamma=0$) and nonmonochromatic ($\gamma=10$) modes.	
    		}
	    \end{figure}

	The correction coefficients (\ref{Eq:CorrectCoeff}) for the scenario with the nonmomochromatic modes in the form of exponential decay are given by
	    \begin{equation}
			C_{k_1 \ldots k_N}=\eta^n \prod\limits_{i=1}^{N}\left[\frac{\sinh\left(\gamma\eta_{k_i-1}/2\right)}{\sinh\left(\gamma/2\right)}\right]^{k_i}.
			\label{Eq:CorCoeff_NM}
		\end{equation}
	In the form of eq.~(\ref{Eq:CorCeff_Simpl}) it can be rewritten as
		    \begin{equation}
				C_{{\{k_{q_1}\ldots k_{q_Q}\}}}=\eta^n \prod\limits_{i=1}^{Q}\left[\frac{\sinh\left(\gamma\eta_{k_{q_i}-1}/2\right)}{\sinh\left(\gamma/2\right)}\right]^{k_{q_i}}.
				\label{Eq:CorCoeff_NM1}
			\end{equation}		
	Their dependence on the dead time is shown in figure~\ref{Fig:GraphTD2}.
	In general, this dependence resembles properties of the same function in the scenario of monochromatic light, cf. figure~\ref{Fig:GraphTD1}.
	However, in the considered case, it vanishes much faster when the dead time increases.

\section{Summary and conclusions}
\label{Sec:Concl}
        
    To summarize, the generalized Hong-Ou-Mandel experiment plays a significant role in understanding nonclassical properties of quantum light.
    It is also used for boson sampling, which is one of the most promising models of nonuniversal quantum computation. 
    Our attention was drawn to bunched photons, which are not usually considered in the context of boson sampling.
    Observations of the corresponding events are usually complicated due to the imperfect abilities of detectors to distinguish between numbers of photons.
    
    We consider detection scenarios for which detection events are solely related to signal photons.
    This means that the rate of dark counts, afterpulses, etc. is negligibly low.
    In this case, the corresponding photocounting probabilities are the probabilities for ideal PNR detectors scaled by the correction coefficients.
    Hence, the formula describing the photon-number distribution in the generalized Hong-Ou-Mandel experiment can be verified even with realistic photon-number resolution.  
    
    Our theoretical consideration is illustrated with examples of two widely-used measurement schemes.
    Firstly, we have dealt with arrays of on/off detectors.
    Secondly, we have considered photocounting with the dead time of detectors.
    For both measurement techniques, the correction coefficients take very small values for the cases with a large number of bunched photons.
    This poses a restriction on the considered method since the corresponding events will be rare. 
    We believe that our results will be useful for experimental verification of the generalized Hong-Ou-Mandel effect and developing validation methods for boson-sampling schemes.

   	V.Ye.L., M.M.B., and A.A.S. acknowledge support from the Department of Physics and Astronomy of the National Academy of Sciences of Ukraine through the project 0120U100857.  
   	V.A.U. acknowledges support from the National Research Foundation of Ukraine through the project 2020.02/0111 ``Nonclassical and hybrid correlations of quantum systems under realistic conditions''.
   	A. A. S. also thanks B. Hage and J. Kr\"oger for enlightening discussions of photodetection with dead time.

\section*{References}
\bibliography{biblio}

\end{document}